\documentclass[conference, letterpaper]{IEEEtran}
\usepackage{cite}
\usepackage[pdftex]{graphicx}
\usepackage[cmex10]{amsmath}
\usepackage{array}
\usepackage{adjustbox}
\usepackage{multirow}
\usepackage{diagbox}
\usepackage{tikz}
\def\checkmark{\tikz\fill[scale=0.4](0,.35) -- (.25,0) -- (1,.7) -- (.25,.15) -- cycle;}
\usepackage{pifont}
\newcommand{\xmark}{\ding{55}}
\usepackage[table, xcdraw]{colortbl}
\newcolumntype{P}[1]{>{\centering\arraybackslash}p{#1}}
\newcolumntype{M}[1]{>{\centering\arraybackslash}m{#1}}

\usepackage{fancyhdr}

\renewcommand{\thispagestyle}[2]{}

\fancypagestyle{plain}{
       \fancyhead{}
       \fancyhead[C]{first page center header}
       \fancyfoot{}
       \fancyfoot[C]{first page center footer}
}
\begin{document}
\title{A Generic Multi-modal Dynamic Gesture Recognition System using Machine Learning}

\author{
\IEEEauthorblockN{Gautham Krishna G\textsuperscript{$\alpha$}, 
Karthik Subramanian Nathan\textsuperscript{$\beta$}, Yogesh Kumar B\textsuperscript{$\gamma$},
Ankith A Prabhu\textsuperscript{$\delta$},\\
Ajay Kannan\textsuperscript{$\pi$}, Vineeth Vijayaraghavan\textsuperscript{$\epsilon$}}

\IEEEauthorblockA{Research Assistant\textsuperscript{$\alpha$}\textsuperscript{$\gamma$}, Undergraduate Student\textsuperscript{$\beta$}\textsuperscript{$\delta$}\textsuperscript{$\pi$}, Director - Research \& Outreach, Solarillion Foundation, Chennai, India\textsuperscript{$\epsilon$}}

\IEEEauthorblockA{Solarillion Foundation\textsuperscript{$\alpha$}\textsuperscript{$\gamma$}\textsuperscript{$\epsilon$}, College of Engineering, Guindy\textsuperscript{$\beta$}\textsuperscript{$\pi$}, SRM University\textsuperscript{$\delta$}}

\IEEEauthorblockA{(gautham.krishna\textsuperscript{$\alpha$}, 
nathankarthik\textsuperscript{$\beta$}, yogesh.bkumar\textsuperscript{$\gamma$}, 
ankithprabhu\textsuperscript{$\delta$}, ajaykannan\textsuperscript{$\pi$}, 
vineethv\textsuperscript{$\epsilon$})@ieee.org}
}

\maketitle

\begin{abstract}

Human Computer Interaction facilitates intelligent communication between humans and computers, in which gesture recognition plays a prominent role. This paper proposes a machine learning system to identify dynamic gestures using tri-axial acceleration data acquired from two public datasets. These datasets - uWave and Sony, were acquired using accelerometers embedded in Wii remotes and smartwatches respectively. A dynamic gesture signed by the user is characterized by a generic set of features extracted across time and frequency domains. The system was analyzed from an end-user perspective and was modelled to operate in three modes. The modes of operation determine the subsets of data to be used for training and testing the system. From an initial set of seven classifiers, three were chosen to evaluate each dataset across all modes rendering the system towards mode-neutrality and dataset-independence. The proposed system is able to classify gestures performed at varying speeds with minimum preprocessing, making it computationally efficient. Moreover, this system was found to run on a low-cost embedded platform - Raspberry Pi Zero (USD 5), making it economically viable.

\end{abstract}


\IEEEpeerreviewmaketitle

\section{Introduction}

Gesture recognition can be defined as the perception of non-verbal communication through an interface that identifies gestures using mathematical, probabilistic and statistical methods. The field of gesture recognition has been experiencing a rapid growth amidst increased interests shown by researchers in the industry. The goal of current research has been the quick and accurate classification of gestures with minimalistic computation, whilst being economically feasible. Gesture recognition can find use in various tasks such as developing aids for the audio-vocally impaired using sign language interpretation, virtual gaming and smart home environments.

Modern gesture recognition systems can be divided into two broad categories - vision based and motion based systems. The vision based system proposed by Chen et al. in \cite{cite:digital} uses digital cameras, and that proposed by Biswas et al. in \cite{cite:xbox} uses infrared  cameras to track the movement of the user. For accurate classification of the gestures, these systems require proper lighting, delicate and expensive hardware and computationally intensive algorithms. On the other hand, motion based systems use data acquired from sensors like accelerometer, gyroscope and flex sensor to identify the gestures being performed by the user. Of late, most gesture recognition systems designed for effective interaction utilize accelerometers for cheaper and accurate data collection. 

Accelerometer-based hand gesture recognition systems deal with either static or dynamic gestures as mentioned in \cite{cite:survey}. Static gestures can be uniquely characterized by identifying their start and end points, while dynamic gestures require the entire data sequence of a gesture sample to be considered. Constructing a dynamic gesture recognition system that is compatible with any user is difficult, as the manner in which the same gesture is performed varies from user-to-user. This variation arises because of the disparate speeds of the dynamic gestures signed by users. To tackle this problem, a common set of features which represent the dynamic nature of the gestures across various users should be selected.  

The authors of this paper propose a gesture recognition system using a generic feature set, implemented on two public datasets using accelerometers - uWave and Sony, as shown in \cite{cite:uwave} and \cite{cite:sony} respectively. This system has been trained and tested across various classifiers and modes, giving equal importance to both accuracy and classification time, unlike most conventional systems. This results in a computationally efficient model for the classification of dynamic gestures which is compatible with low-cost systems.

The rest of the paper is organized as follows. Section \ref{section:related_work} presents the related work in the area of Gesture Recognition. Section \ref{section:problem_statement} states about the problem statement of the paper. Sections \ref{section:datasets} \& \ref{section:dataset_preprocessing} deal with the datasets and pre-processing used in building the model. Section \ref{section:feature_extraction} showcases the features extracted from the datasets. Section \ref{section:mode} discusses about the different modes provided to the end-user. Section \ref{section:experiment} describes the model used and the experiments performed. Section \ref{section:results} enumerates the results analyzed in the paper. Section \ref{section:conclusion} finally concludes the paper.

\section{Related Work}
\label{section:related_work}

Since the inception of gesture recognition systems, there has been a plethora of research in this domain using accelerometers. Specifically tri-axial accelerometers have been in the spotlight recently owing to their low-cost and low-power requirements in conjunction with their miniature sizes, making them ideal for embedding into the consumer electronic platform. Previous gesture recognition systems have also used sensors such as flex sensors and gyroscope, but they have their own shortcomings. The glove based system proposed by Zimmerman et al. in \cite{cite:flex} utilizes flex sensors, which requires intensive calibration on all sensors. The use of the these sensors increases the cost of the system and also makes the system physically cumbersome. These shortcomings make inertial sensors like accelerometers and gyroscope, a better alternative. In this paper, datasets employing accelerometers were preferred over gyroscopes, as processing the data from gyroscopes results in a higher computational burden.

Contemporary gesture recognition systems employ Dynamic Time Warping (DTW) algorithms for classification of gestures. For each user, Liu et al. \cite{cite:uwave} employs DTW to compute the look-up table (template) for each gesture, but it is not representative of all users in the dataset, thereby not generalizing for the user-independent paradigm. To achieve a generic look-up table for each gesture that represents multiple users, the proposed system in \cite{cite:bangladesh} uses the concept of idealizing a template wherein, the gestures exhibiting the least cost when internal DTW is performed is chosen as the look-up table. Furthermore, DTW is performed again for gesture classification while testing, making the model computationally very expensive to be used in a low-cost embedded platform. The accelerometer-based gesture recognition system proposed in \cite{cite:hmm} uses continuous Hidden Markov Models (HMMs), but their computational complexity is commensurate with the size of the feature vectors which increase rapidly. In addition, choosing the optimum number of states is difficult in multi-user temporal sequences, thereby increasing the complexity of estimating probability functions. 

Moreover, the length of the time series acceleration values of a gesture sample is made equal and quantization is performed in the system proposed in \cite{cite:uwave}. This makes the data points either lossy or redundant. However, the paper proposed utilizes the data points as provided in the datasets without any windowing or alteration, thereby decreasing the computational costs whilst not compromising on efficiency.

In the system employed in \cite{cite:feature_neurofuzzy}, Helmi et al. have selected a set of features that has been implemented on a dataset that contains gestures performed by a single user, making them gesture-dependent and not taking into account the concept of generic features encompassing multiple users. The need for a generic set of features which capture the similarities between gestures, motivated the authors of this paper to implement a feature set that can be applied to any system.

\section{Problem Statement}
\label{section:problem_statement}

Previous works in haptic-based gesture recognition systems employed computationally expensive algorithms for identification of gestures with instrumented and multifarious sensors, making them reliant on specialized hardware. Moreover, most gesture recognition systems extract features that are model-dependent, and fail to provide the user a choice between accuracy and classification time. Thus a need for a flexible gesture recognition system that identifies dynamic gestures accurately with minimalistic hardware, along with a generic feature set arises.

To overcome this problem, this paper presents a machine-learning based dynamic gesture recognition system that utilizes accelerometers along with a generic set of features that can be implemented across any model with adequate gesture samples. The system provides the end-user the option to choose between accuracy and classification time, thereby giving equal importance to both.

\section{Datasets}
\label{section:datasets}

\begin{table}[ht]
\caption{Datasets characterized by their respective attributes}
\label{table:dataset}
\begin{adjustbox}{width=0.481\textwidth, height=0.028\textwidth}
\begin{tabular}{|c|c|c|c|c|c|}
\hline
Dataset			  &   Users ($U$) &   Gestures ($N_G$) &   Samples per Gesture ($S_G$) &   Days ($N_D$) &   $N_G{}_S$ \\
\hline
uWave ($D_u$)     &       8 &          8 &          10 &      7 & 4480 \\
\hline
Sony ($D_S$)      &       8 &         20 &          20 &      - & 3200 \\
\hline
\end{tabular}
\end{adjustbox}
\end{table}

For this paper, the authors have chosen two public gesture datasets, viz. uWave dataset ($D_u$) and Sony dataset ($D_S$). The datasets were selected owing to their large user campaign of 8 users, with a multitude of gesture samples for a variety of gestures. $D_u$ encompasses an 8-gesture vocabulary with 560 gesture samples per dataset over a period of 7 days, while $D_S$ consists of a collection of 20 gestures with 160 gesture samples each. This shows the diverse nature of the datasets. Both the datasets are characterized by $U$ users signing $N_G$ gestures with $S_G$ samples per gesture, over $N_D$ days with the total number of gesture samples being $N_G{}_S$ per dataset, as shown in Table \ref{table:dataset}.

\begin{figure}[ht]
\centering
\includegraphics[width=120pt, height=100pt]{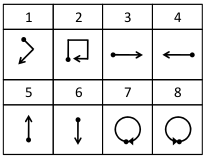}
\caption{uWave Gestures}
\label{fig:uwave_gestures}
\end{figure}

$D_u$ comprises of 3-D accelerations (g-values) that were recorded using a Wii Remote. The start of a gesture is indicated by pressing the 'A' button on the Wii Remote and the end is detected by releasing the button as mentioned in \cite{cite:uwave}. $D_u$ consists of four gestures which have 1-D motion while the remaining gestures have 2-D motion, as depicted in Figure \ref{fig:uwave_gestures}.

\begin{figure}[ht]
\centering
\includegraphics[width=150pt, height=170pt]{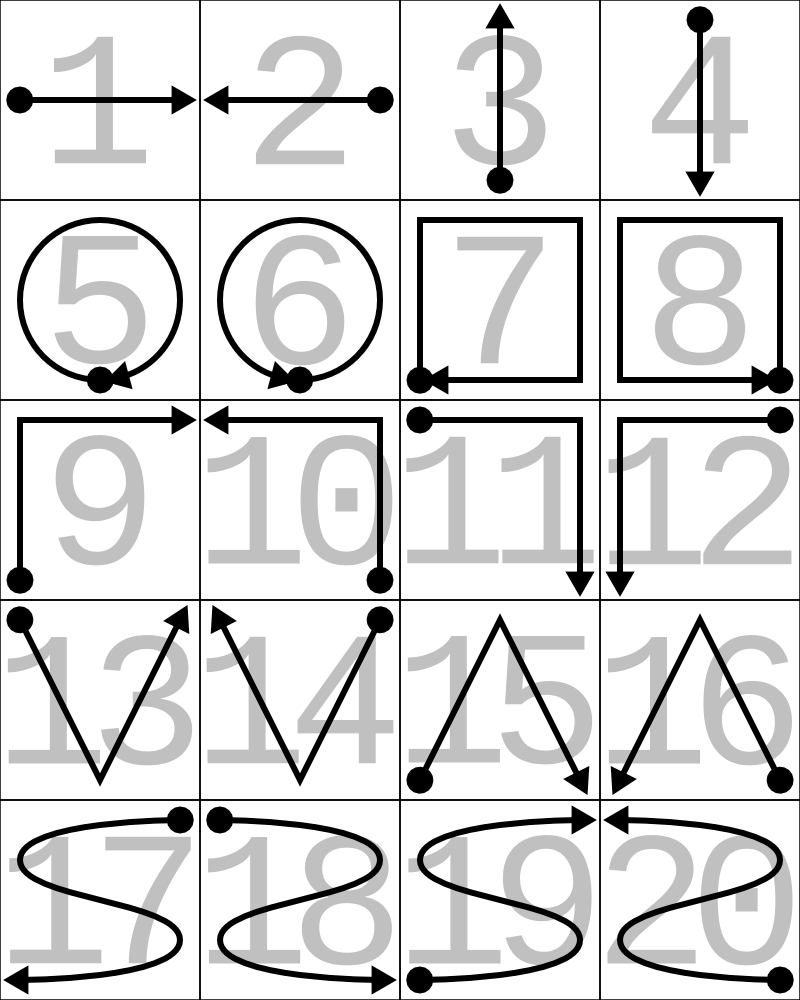}
\caption{Sony Gestures}
\label{fig:sony_gestures}
\end{figure}

\begin{table*}[ht]
\centering
\caption{Confusion Matrix for $U_M$ mode on $D_S$; Brown: Gesture Signed, Yellow: Gesture Classified, Green: Correct Classifications, Red: Incorrect Classifications}
\label{table:sony_mixed_confmat}
\begin{tabular}{|c|c|c|c|c|c|c|c|c|c|c|c|c|c|c|c|c|c|c|c|c|}
\hline
\rowcolor[HTML]{FFFE65}
{\color[HTML]{000000} }    & 1                             & 2                         & 3                             & 4                             & 5                             & 6                         & 7                             & 8                         & 9                            & 10                        & 11                            & 12                            & 13                            & 14                            & 15                            & 16                            & 17                        & 18                        & 19                        & 20                        \\ \hline
\cellcolor[HTML]{EDB13D}1  & \cellcolor[HTML]{67FD9A}100     & 0                         & 0                             & 0                             & 0                             & 0                         & 0                             & 0                         & 0                            & 0                         & 0                             & 0                             & 0                             & 0                             & 0                             & 0                             & 0                         & 0                         & 0                         & 0                         \\ \hline
\cellcolor[HTML]{EDB13D}2  & 0                             & \cellcolor[HTML]{67FD9A}100 & 0                             & 0                             & 0                             & 0                         & 0                             & 0                         & 0                            & 0                         & 0                             & 0                             & 0                             & 0                             & 0                             & 0                             & 0                         & 0                         & 0                         & 0                         \\ \hline
\cellcolor[HTML]{EDB13D}3  & 0                             & 0                         & \cellcolor[HTML]{67FD9A}100     & 0                             & 0                             & 0                         & 0                             & 0                         & 0                            & 0                         & 0                             & 0                             & 0                             & 0                             & 0                             & 0                             & 0                         & 0                         & 0                         & 0                         \\ \hline
\cellcolor[HTML]{EDB13D}4  & \cellcolor[HTML]{F8524D}2.5 & 0                         & 0                             & \cellcolor[HTML]{67FD9A}97.5 & 0                             & 0                         & 0                             & 0                         & 0                            & 0                         & 0                             & 0                             & 0                             & 0                             & 0                             & 0                             & 0                         & 0                         & 0                         & 0                         \\ \hline
\cellcolor[HTML]{EDB13D}5  & 0                             & 0                         & 0                             & 0                             & \cellcolor[HTML]{67FD9A}100     & 0                         & 0                             & 0                         & 0                            & 0                         & 0                             & 0                             & 0                             & 0                             & 0                             & 0                             & 0                         & 0                         & 0                         & 0                         \\ \hline
\cellcolor[HTML]{EDB13D}6  & 0                             & 0                         & 0                             & 0                             & 0                             & \cellcolor[HTML]{67FD9A}100 & 0                             & 0                         & 0                            & 0                         & 0                             & 0                             & 0                             & 0                             & 0                             & 0                             & 0                         & 0                         & 0                         & 0                         \\ \hline
\cellcolor[HTML]{EDB13D}7  & 0                             & 0                         & 0                             & 0                             & 0                             & 0                         & \cellcolor[HTML]{67FD9A}97.5 & 0                         & 0                            & 0                         & 0                             & 0                             & 0                             & \cellcolor[HTML]{F8524D}2.5 & 0                             & 0                             & 0                         & 0                         & 0                         & 0                         \\ \hline
\cellcolor[HTML]{EDB13D}8  & 0                             & 0                         & 0                             & 0                             & 0                             & 0                         & 0                             & \cellcolor[HTML]{67FD9A}100 & 0                            & 0                         & 0                             & 0                             & 0                             & 0                             & 0                             & 0                             & 0                         & 0                         & 0                         & 0                         \\ \hline
\cellcolor[HTML]{EDB13D}9  & 0                             & 0                         & 0                             & 0                             & 0                             & 0                         & 0                             & 0                         & \cellcolor[HTML]{67FD9A}95 &  \cellcolor[HTML]{F8524D}5                      & 0                             & 0                             & 0                             & 0                             & 0                             & 0                             & 0                         & 0                         & 0                         & 0                         \\ \hline
\cellcolor[HTML]{EDB13D}10 & 0                             & 0                         & 0                             & 0                             & 0                             & 0                         & 0                             & 0                         & 0                            & \cellcolor[HTML]{67FD9A}100 & 0                             & 0                             & 0                             & 0                             & 0                             & 0                             & 0                         & 0                         & 0                         & 0                         \\ \hline
\cellcolor[HTML]{EDB13D}11 & 0                             & 0                         & 0                             & 0                             & 0                             & 0                         & 0                             & 0                         & 0                            & 0                         & \cellcolor[HTML]{67FD9A}100     & 0                             & 0                             & 0                             & 0                             & 0                             & 0                         & 0                         & 0                         & 0                         \\ \hline
\cellcolor[HTML]{EDB13D}12 & 0                             & 0                         & 0                             & 0                             & 0                             & 0                         & 0                             & 0                         & 0                            & 0                         & \cellcolor[HTML]{F8524D}2.5 & \cellcolor[HTML]{67FD9A}97.5 & 0                             & 0                             & 0                             & 0                             & 0                         & 0                         & 0                         & 0                         \\ \hline
\cellcolor[HTML]{EDB13D}13 & 0                             & 0                         & 0                             & 0                             & 0                             & 0                         & 0                             & 0                         & 0                            & 0                         & 0                             & 0                             & \cellcolor[HTML]{67FD9A}97.5 & \cellcolor[HTML]{F8524D}2.5 & 0                             & 0                             & 0                         & 0                         & 0                         & 0                         \\ \hline
\cellcolor[HTML]{EDB13D}14 & 0                             & 0                         & 0                             & 0                             & 0                             & 0                         & 0                             & 0                         & 0                            & 0                         & 0                             & 0                             & 0                             & \cellcolor[HTML]{67FD9A}97.5 & \cellcolor[HTML]{F8524D}2.5 & 0                             & 0                         & 0                         & 0                         & 0                         \\ \hline
\cellcolor[HTML]{EDB13D}15 & 0                             & 0                         & 0                             & 0                             & 0                             & 0                         & 0                             & 0                         & 0                            & 0                         & 0                             & 0                             & 0                             & 0                             & \cellcolor[HTML]{67FD9A}97.5 & \cellcolor[HTML]{F8524D}2.5 & 0                         & 0                         & 0                         & 0                         \\ \hline
\cellcolor[HTML]{EDB13D}16 & 0                             & 0                         & \cellcolor[HTML]{F8524D}2.5 & 0                             & \cellcolor[HTML]{F8524D}2.5 & 0                         & 0                             & 0                         & 0                            & 0                         & 0                             & 0                             & 0                             & 0                             & \cellcolor[HTML]{F8524D}2.5 & \cellcolor[HTML]{67FD9A}92.5 & 0                         & 0                         & 0                         & 0                         \\ \hline
\cellcolor[HTML]{EDB13D}17 & 0                             & 0                         & 0                             & 0                             & 0                             & 0                         & 0                             & 0                         & 0                            & 0                         & 0                             & 0                             & 0                             & 0                             & 0                             & 0                             & \cellcolor[HTML]{67FD9A}100 & 0                         & 0                         & 0                         \\ \hline
\cellcolor[HTML]{EDB13D}18 & 0                             & 0                         & 0                             & 0                             & 0                             & 0                         & 0                             & 0                         & 0                            & 0                         & 0                             & 0                             & 0                             & 0                             & 0                             & 0                             & 0                         & \cellcolor[HTML]{67FD9A}100 & 0                         & 0                         \\ \hline
\cellcolor[HTML]{EDB13D}19 & 0                             & 0                         & 0                             & 0                             & 0                             & 0                         & 0                             & 0                         & 0                            & 0                         & 0                             & 0                             & 0                             & 0                             & 0                             & 0                             & 0                         & 0                         & \cellcolor[HTML]{67FD9A}100 & 0                         \\ \hline
\cellcolor[HTML]{EDB13D}20 & 0                             & 0                         & 0                             & 0                             & 0                             & 0                         & 0                             & 0                         & 0                            & 0                         & 0                             & 0                             & 0                             & 0                             & 0                             & 0                             & 0                         & 0                         & 0                         & \cellcolor[HTML]{67FD9A}100 \\ \hline
\end{tabular}
\end{table*}

$D_S$ was recorded using a tri-axial accelerometer of a first generation Sony Smartwatch which was worn on the right wrist of the user. Each gesture instance was performed by tapping the smartwatch screen to indicate the start and end of the gesture. The data recorded from the smartwatch consists of timestamps from different clock sources of an Android device, along with the acceleration (g-values) measured across each axis, as mentioned in \cite{cite:sony}. In $D_S$, there are four gestures which have motions in 1-D while the remaining gestures have motions in 2-D as illustrated in Figure \ref{fig:sony_gestures}.

\section{Dataset Preprocessing}
\label{section:dataset_preprocessing}

To generalize the datasets, we eliminate the dependency on time from the datasets, which consist of  timestamps along with the raw g-values. The set of g-values per gesture sample ($\gamma$), each with a cardinality of $n_\gamma$ is given by,

\begin{equation}
\begin{split}
\label{equation:gamma}
\gamma = \{g_x^i, \ g_y^i, \ g_z^i\}, \forall \ i \in [1, \ n_\gamma] \ \\ where,\  g_x^i , \ g_y^i ,  \ g_z^i \ are \\ g-values \ of \ the \ accelerometer
\end{split}
\end{equation}

On account of the dynamic nature and varying speeds at which different gestures are signed by the users, $n_\gamma$ varies between different gesture samples. From equation \ref{equation:gamma}, an overall dataset ($G$) can be defined by,

\begin{equation}
\label{equation:dataset}
G = \bigcup_{i=1}^{N_G{}_S}\gamma^i
\end{equation}

Equation \ref{equation:dataset} is representative of both $D_S$ and $D_u$ datasets. No preprocessing other than elimination of timestamps was done to both the datasets as the g-values in $\gamma$ would be altered, thereby making the datasets lossy.

\section{Feature Extraction}
\label{section:feature_extraction}

\subsection{Feature Characterization}
The proposed system uses the features mentioned in Table \ref{table:features}, which have been already utilized in previous accelerometer-based studies. From \cite{cite:activity}, the significance of Minimum, Maximum, Mean, Skew, Kurtosis and Cross correlation have been shown for Activity Recognition, which is a superset of Gesture Recognition. The inter-axial Pearson product-moment (PM) correlation coefficients as a feature has been signified in \cite{cite:pearson}. The Spectral  Energy as a feature has been illustrated in \cite{cite:fft_features}. The aforementioned features were iteratively eliminated in various domains until the efficiency of the model was the highest.

\begin{table}[ht]
\centering
\caption{Feature Characterization in Time and Frequency Domains}
\label{table:features}
\begin{tabular}{|l|l|l|l|}
\hline
\multicolumn{1}{|c|}{\multirow{2}{*}{Features$\backslash	$Domain}} & \multicolumn{1}{c|}{\multirow{2}{*}{Time}} & \multicolumn{2}{c|}{\multirow{1}{*}{Frequency}} \\ \cline{3-4}
\multicolumn{1}{|c|}{}                                 & \multicolumn{1}{c|}{}                      &  \multicolumn{1}{c|}{FFT}            & \multicolumn{1}{c|}{HT}            \\ \hline
Mean                                                   &                                           \checkmark ($T^1$)&  \ \ \ \ \xmark    &   \checkmark  ($H^1$)          \\ \hline
Skew                                                   &                                           \checkmark ($T^2$)&   \ \ \ \ \xmark   &   \checkmark ($H^2$)           \\ \hline
Kurtosis                                               &                                           \checkmark ($T^3$)&   \ \ \ \ \xmark   &  \ \ \ \ \xmark                \\ \hline
PM correlation coefficients                             &                                           \checkmark ($T^4$)&   \ \ \ \ \xmark   &   \ \ \ \ \xmark               \\ \hline
Cross correlation                                     &                                           \checkmark ($T^5$)&   \ \ \ \ \xmark   &   \ \ \ \ \xmark               \\ \hline
Energy                                                 &  \ \ \ \ \xmark                             &   \checkmark  ($F^1$)                &  \checkmark  ($H^3$)           \\ \hline
Minimum                                                &   \ \ \ \ \xmark                                         &  \ \ \ \ \xmark                      & \checkmark ($H^4$)             \\ \hline
Maximum                                                &   \ \ \ \ \xmark                                         &   \ \ \ \ \xmark                     &   \checkmark ($H^5$)           \\ \hline
\end{tabular}
\end{table}

\begin{table*}
\centering
\caption{Confusion Matrix for $U_I$ mode on $D_S$; Brown: Gesture Signed, Yellow: Gesture Classified, Green: Correct Classifications, Red: Incorrect Classifications}
\label{table:sony_independent_confmat}
\begin{tabular}{|c|c|c|c|c|c|c|c|c|c|c|c|c|c|c|c|c|c|c|c|c|}
\hline
\rowcolor[HTML]{FFFE65}
{\color[HTML]{000000} }    & 1                            & 2                         & 3                            & 4                         & 5                            & 6                            & 7                            & 8                           & 9                            & 10                           & 11                           & 12                           & 13                           & 14                        & 15                           & 16                          & 17                           & 18                          & 19                          & 20                           \\ \hline
\cellcolor[HTML]{EDB13D}1  & \cellcolor[HTML]{67FD9A}100    & 0                         & 0                            & 0                         & 0                            & 0                            & 0                            & 0                           & 0                            & 0                            & 0                            & 0                            & 0                            & 0                         & 0                            & 0                           & 0                            & 0                           & 0                           & 0                            \\ \hline
\cellcolor[HTML]{EDB13D}2  & 0                            & \cellcolor[HTML]{67FD9A}100 & 0                            & 0                         & 0                            & 0                            & 0                            & 0                           & 0                            & 0                            & 0                            & 0                            & 0                            & 0                         & 0                            & 0                           & 0                            & 0                           & 0                           & 0                            \\ \hline
\cellcolor[HTML]{EDB13D}3  & 0                            & 0                         & \cellcolor[HTML]{67FD9A}95 & 0                         & 0                            & 0                            & 0                            & 0                           & \cellcolor[HTML]{F8524D}5 & 0                            & 0                            & 0                            & 0                            & 0                         & 0                            & 0                           & 0                            & 0                           & 0                           & 0                            \\ \hline
\cellcolor[HTML]{EDB13D}4  & 0                            & 0                         & 0                            & \cellcolor[HTML]{67FD9A}100 & 0                            & 0                            & 0                            & 0                           & 0                            & 0                            & 0                            & 0                            & 0                            & 0                         & 0                            & 0                           & 0                            & 0                           & 0                           & 0                            \\ \hline
\cellcolor[HTML]{EDB13D}5  & 0                            & 0                         & 0                            & 0                         & \cellcolor[HTML]{67FD9A}65 & \cellcolor[HTML]{F8524D}15 & 0                            & 0                           & \cellcolor[HTML]{F8524D}5 & 0                            & 0                            & 0                            & 0                            & 0                         & \cellcolor[HTML]{F8524D}15 & 0                           & 0                            & 0                           & 0                           & 0                            \\ \hline
\cellcolor[HTML]{EDB13D}6  & 0                            & 0                         & 0                            & 0                         & 0                            & \cellcolor[HTML]{67FD9A}95 & \cellcolor[HTML]{F8524D}5 & 0                           & 0                            & 0                            & 0                            & 0                            & 0                            & 0                         & 0                            & 0                           & 0                            & 0                           & 0                           & 0                            \\ \hline
\cellcolor[HTML]{EDB13D}7  & 0                            & 0                         & 0                            & 0                         & 0                            & 0                            & \cellcolor[HTML]{67FD9A}90  & \cellcolor[HTML]{F8524D}10 & 0                            & 0                            & 0                            & 0                            & 0                            & 0                         & 0                            & 0                           & 0                            & 0                           & 0                           & 0                            \\ \hline
\cellcolor[HTML]{EDB13D}8  & 0                            & 0                         & 0                            & 0                         & 0                            & 0                            & 0                            & \cellcolor[HTML]{67FD9A}100   & 0                            & 0                            & 0                            & 0                            & 0                            & 0                         & 0                            & 0                           & 0                            & 0                           & 0                           & 0                            \\ \hline
\cellcolor[HTML]{EDB13D}9  & \cellcolor[HTML]{F8524D}5 & 0                         & \cellcolor[HTML]{F8524D}5 & 0                         & 0                            & 0                            & 0                            & 0                           & \cellcolor[HTML]{67FD9A}85 & \cellcolor[HTML]{F8524D}5 & 0                            & 0                            & 0                            & 0                         & 0                            & 0                           & 0                            & 0                           & 0                           & 0                            \\ \hline
\cellcolor[HTML]{EDB13D}10 & 0                            & 0                         & 0                            & 0                         & 0                            & 0                            & 0                            & 0                           & \cellcolor[HTML]{F8524D}10  & \cellcolor[HTML]{67FD9A}90  & 0                            & 0                            & 0                            & 0                         & 0                            & 0                           & 0                            & 0                           & 0                           & 0                            \\ \hline
\cellcolor[HTML]{EDB13D}11 & 0                            & 0                         & 0                            & 0                         & 0                            & 0                            & 0                            & 0                           & 0                            & 0                            & \cellcolor[HTML]{67FD9A}95 & \cellcolor[HTML]{F8524D}5 & 0                            & 0                         & 0                            & 0                           & 0                            & 0                           & 0                           & 0                            \\ \hline
\cellcolor[HTML]{EDB13D}12 & 0                            & 0                         & 0                            & 0                         & 0                            & 0                             & 0                            & 0                           & 0                            & 0                            & 0                            & \cellcolor[HTML]{67FD9A}100    & 0                            & 0                         & 0                            & 0                           & 0                            & 0                           & 0                           & 0                            \\ \hline
\cellcolor[HTML]{EDB13D}13 & 0                            & 0                         & 0                            & 0                         & 0                            & 0                            & 0                            & 0                           & 0                            & 0                            & 0                            & \cellcolor[HTML]{F8524D}5 & \cellcolor[HTML]{67FD9A}95 & 0                         & 0                            & 0                           & 0                            & 0                           & 0                           & 0                            \\ \hline
\cellcolor[HTML]{EDB13D}14 & 0                            & 0                         & 0                            & 0                         & 0                            & 0                            & 0                            & 0                           & 0                            & 0                            & 0                            & 0                            & 0                            & \cellcolor[HTML]{67FD9A}100 & 0                            & 0                           & 0                            & 0                           & 0                           & 0                            \\ \hline
\cellcolor[HTML]{EDB13D}15 & 0                            & 0                         & 0                            & 0                         & 0                            & 0                            & 0                            & 0                           & 0                            & 0                            & 0                            & 0                            & 0                            & 0                         & \cellcolor[HTML]{67FD9A}90  & \cellcolor[HTML]{F8524D}10 & 0                            & 0                           & 0                           & 0                            \\ \hline
\cellcolor[HTML]{EDB13D}16 & 0                            & 0                         & 0                            & 0                         & 0                            & 0                            & 0                            & 0                           & 0                            & 0                            & 0                            & 0                            & 0                            & 0                         & \cellcolor[HTML]{F8524D}20  & \cellcolor[HTML]{67FD9A}80 & 0                            & 0                           & 0                           & 0                            \\ \hline
\cellcolor[HTML]{EDB13D}17 & 0                            & 0                         & 0                            & 0                         & 0                            & 0                            & 0                            & 0                           & 0                            & 0                            & 0                            & 0                            & 0                            & 0                         & 0                            & 0                           & \cellcolor[HTML]{67FD9A}95 & 0                           & 0                           & \cellcolor[HTML]{F8524D}5 \\ \hline
\cellcolor[HTML]{EDB13D}18 & 0                            & 0                         & 0                            & 0                         & \cellcolor[HTML]{F8524D}5 & 0                            & 0                            & 0                           & 0                            & 0                            & 0                            & 0                            & 0                            & 0                         & 0                            & 0                           & \cellcolor[HTML]{F8524D}5 & \cellcolor[HTML]{67FD9A}90 & 0                           & 0                            \\ \hline
\cellcolor[HTML]{EDB13D}19 & 0                            & 0                         & 0                            & 0                         & 0                            & 0                            & 0                            & 0                           & 0                            & 0                            & 0                            & 0                            & 0                            & 0                         & 0                            & 0                           & 0                            & 0                           & \cellcolor[HTML]{67FD9A}100   & 0                            \\ \hline
\cellcolor[HTML]{EDB13D}20 & 0                            & 0                         & 0                            & 0                         & 0                            & 0                            & 0                            & 0                           & 0                            & 0                            & 0                            & 0                            & 0                            & 0                         & 0                            & 0                           & 0                            & 0                           & \cellcolor[HTML]{F8524D}30 & \cellcolor[HTML]{67FD9A}70  \\ \hline
\end{tabular}
\end{table*}

\subsection{Domain Characterization}
The set of extracted features were then applied in both time and frequency domains. The features that were extracted in the time domain are Mean, Skew, Kurtosis, PM correlation coefficients and Cross correlation. These features along each axis of the gesture samples of a dataset ($G$) in time domain constitute the $TF$ vector, as given in Equation \ref{equation:time_features}.

\begin{equation}
\label{equation:time_features}
TF = \bigcup_{j=1}^{5}\{T_x^j, \ T_y^j, \ T_z^j\}
\end{equation}

The time-domain sequences were converted to frequency domain using Fast Fourier Transform ($FFT$) and only Energy, as mentioned in \cite{cite:fft_features} was used, while the other features from $FFT$ did not provide any significant improvements. The feature vector ($FF$) along each axis of the gesture samples of a dataset ($G$) is represented by Equation \ref{equation:fft_features}.

\begin{equation}
\label{equation:fft_features}
FF = \{F_x^1, \ F_y^1, \ F_z^1\}
\end{equation}

From \cite{cite:hilbert_handwriting}, it can be seen that the Hilbert Transform ($HT$) hypothesized as a feature was successful in providing a competitive recognition rate for camera-based handwriting gestures. This novel approach was used in the accelerometer-based gesture recognition model proposed by the authors.

Hilbert transform is a linear transformation used in signal processing, that accepts as input a temporal signal, and produces an analytic signal. The analytic signal consists of a real and an imaginary part as explained in \cite{cite:hilbert_handwriting}, where the negative half of the frequency spectrum is zeroed out. The real part represents the input signal and the imaginary part ($y$) is representative of the Hilbert transformed signal, which is phase shifted by $\pm90^{\circ}$. For an input signal ($x(t)$), the analytic signal ($x_a(t)$) after Hilbert transform is given by Equation \ref{equation:hilbert}.

\begin{equation}
\begin{split}
\label{equation:hilbert}
x_a = F^{-1}(F(x)2U) = x+iy, \ \\ where, \ F - Fourier \ transform, \\ U - \ Unit \ step \ function, \\ y  - Hilbert \ transform \ of \ x.
\end{split}
\end{equation}

This approach of using Hilbert Transform for feature extraction is applied across all gesture samples, and the features - Mean, Skew, Minimum, Maximum and Energy are calculated as shown in Table \ref{table:features}. The features Mean and Skew, which have been used in $TF$, and Energy which is calculated in $FF$ are also used as Hilbert features since reusing them here yields a better performance for classifying different gestures. These Hilbert transformed features along each axis on a dataset $(G$) make up the $HF$ vector, which is given by Equation \ref{equation:hilbert_features}.

\begin{equation}
\label{equation:hilbert_features}
HF = \bigcup_{j=1}^{5}\{H_x^j, \ H_y^j, \ H_z^j\}
\end{equation}

The FeatureSet ($FS$) for a single dataset ($G$) is formed from Equations \ref{equation:time_features}, \ref{equation:fft_features} and \ref{equation:hilbert_features} by appending all the feature vectors - $TF$, $FF$ and $HF$ calculated across a dataset ({$G$}), as shown in Equation \ref{equation:featureset}.

\begin{equation}
\label{equation:featureset}
FS = \bigcup_{j=1}^{N_G{}_S}\{TF \cup FF \cup HF\}^j
\end{equation}

\section{End-User Modelling}
\label{section:mode}

This paper enables the end-user to select any one of three proposed modes of operation - User Dependent, Mixed User and User Independent. The User Dependent ($U_D$) mode is an estimator of how well the system performs when the train-test split is between the gestures of a single user. Mixed User ($U_M$) is representative of the complete set of gestures of all participants. The User Independent ($U_I$) mode employs a stratified k-fold cross validation technique which corresponds to training on a number of users and testing on the rest.

\begin{table*}[ht]
\centering
\caption{Average Accuracies (Acc) for the datasets $D_u$ and $D_S$ and Time Taken (in seconds) for classification of a single gesture sample across all modes; Green: Highest Efficiencies in all modes, Yellow: Least Classification Times taken for a gesture sample in all modes}
\label{table:eff_and_time}
\begin{tabular}{|c|c|c|c|c|c|c|c|c|c|c|c|c|}
\hline
      & \multicolumn{6}{c|}{uWave ($D_u$)}                                                                               & \multicolumn{6}{c|}{Sony ($D_s$)}                                                                                \\ \hline
Classifier$\backslash$Mode  & \multicolumn{2}{c|}{User Dependent ($U_D$)} & \multicolumn{2}{c|}{Mixed User ($U_M$)} & \multicolumn{2}{c|}{User Independent ($U_D$)} & \multicolumn{2}{c|}{User Dependent ($U_D$)} & \multicolumn{2}{c|}{Mixed User ($U_M$)} & \multicolumn{2}{c|}{User Independent ($U_D$)} \\ \hline
             & Acc          & Time         & Acc      & Time         & Acc     & Time            & Acc          & Time         & Acc          & Time      & Acc         & Time           \\ \hline
Extra Trees  & \cellcolor[HTML]{67FD9A}97.76          & 0.6287       & \cellcolor[HTML]{67FD9A}97.85           & 0.6873      & \cellcolor[HTML]{67FD9A}82.49               & 0.6853            & \cellcolor[HTML]{67FD9A}95.88         & 0.6038       & \cellcolor[HTML]{67FD9A}98.63          & 0.6538      & \cellcolor[HTML]{67FD9A}75.1               & 0.669          \\ \hline
Random Forest       & 97.41          & 0.6991       & 95.45           & 0.73        & 77.91               & 0.6995            & 95.25          & 0.6887       & 97.13           & 0.7041      & 70.13               & 0.686          \\ \hline
Gradient Boosting   & 93.75          & 0.0072       & 94.38           & 0.0076      & 75.64               & 0.0078            & 90.5           & 0.0055       & 95.5            & 0.0057      & 66.41                & 0.0057         \\ \hline
Bagging             & 92.74          & 0.1526       & 94.19           & 0.1527      & 76.64               & 0.1528            & 93.5           & 0.1673       & 93.37           & 0.1527      & 62.44                 & 0.1529         \\ \hline
Decision Trees      & 89.55          & 0.0015       & 84.11           & 0.0053      & 66.73               & 0.0015            & 84.13          & 0.0014       & 86.25           & 0.0051      & 50.9                 & 0.0015         \\ \hline
Naive Bayes         & 91.16          & 0.0273       & 71.96           & 0.0292      & 64.66               & 0.0273            & 91.38          & 0.0118       & 65.5            & 0.0116      & 54.35                 & 0.0117         \\ \hline
Ridge Classifier    & 97.5           & \cellcolor[HTML]{FFFE65}0.0013       & 83.84           &\cellcolor[HTML]{FFFE65}0.0013      & 74.64               & \cellcolor[HTML]{FFFE65}0.0013            & 94.13         & \cellcolor[HTML]{FFFE65}0.0013       & 77.75           & \cellcolor[HTML]{FFFE65}0.0014      & 61.59                 & \cellcolor[HTML]{FFFE65}0.0013         \\ \hline
\end{tabular}
\end{table*}

\section{Experiment}
\label{section:experiment}

The three modes $U_D$, $U_M$ and $U_I$, as explained in Section \ref{section:mode}, were first trained and tested on an Intel Core i5 CPU @ 2.20 GHz, operating on Ubuntu 16.04. The proposed system was initially implemented using seven classification algorithms that were identified from \cite{cite:gradient}, \cite{cite:ridge} and \cite{cite:bagging} - Extremely Randomized Trees (Extra Trees), Random Forests, Gradient Boosting, Bagging, Decision Trees, Naive Bayes and Ridge Classifier.

Upon further analysis, the seven classifiers were found to run on a low-cost Raspberry Pi Zero, and the accuracies and time taken for a gesture sample to be classified are catalogued for the three modes across both datasets, as shown in Table \ref{table:eff_and_time}. From this set of seven classifiers, the Extra Trees ($ET$), Gradient Boosting ($GB$) and Ridge Classifier ($RC$) were chosen, based on the inferences from Section \ref{subsection:results_A}. Each of the three modes is evaluated using the three classifiers individually, and the efficiencies are noted and analyzed for both the datasets $D_u$ and $D_S$ in Sections \ref{subsection:results_B} and \ref{subsection:results_C} respectively.

\section{Results}
\label{section:results}

\subsection{Evaluation of Classifiers}
\label{subsection:results_A}

The classifiers chosen in Section \ref{section:experiment} were further analyzed for each dataset based on their computational characteristics. It can be inferred that $ET$ yields the highest accuracy in all modes for both datasets $D_u$ and $D_S$, among all the three chosen classifiers. Apart from being more accurate than Random Forest, $ET$ is also computationally less expensive as stated in \cite{cite:EToverRT}. It can also observed that the time taken for classification of a gesture sample is the least in $RC$. $GB$, whilst yielding a significantly lower classification time than $ET$, provides reasonable accuracies and hence was chosen.

Based on the users' preferences, a trade-off can be made between efficiency and classification time of a gesture sample among the three classifiers. To evaluate the performance of this model in all the modes across both datasets, evaluation metrics - cross-validation scores and confusion matrices were taken for the results obtained from the $ET$ classifier, as it has the highest efficiency and an acceptable classification time.

\begin{table}[ht]
\caption{Accuracies of 8 users for both datasets $D_u$ and $D_S$ in User Dependent mode ($U_D$) using Extra Trees Classifier}
\label{table:user_dependent}
\centering
\begin{adjustbox}{width=0.481\textwidth, height=0.028\textwidth}
\begin{tabular}{|c|c|c|c|c|c|c|c|c|c|}
\hline
Data$\backslash$User & 1     & 2    & 3     & 4     & 5     & 6     & 7      & 8   & Avg       \\ \hline
$D_u$     & 96.42 & 93.5 & 98.57 & 97.14 & 99.28 & 99.28 & 97.88  & 100 & 97.76 \\ \hline
$D_S$     & 98    & 92   & 94    & 93    & 98    & 100   & 95     & 97  & 95.88    \\ \hline
\end{tabular}
\end{adjustbox}
\end{table}

\subsection{uWave Dataset Analysis}
\label{subsection:results_B}

In the $U_D$ mode on $D_u$, each user's data was divided into a randomized 75\%-25\% train-test split across all 8 gestures. The average recognition rate achieved over all 8 users was found to be 97.76\%, using $ET$. The individual users' efficiencies for the same can be observed in Table \ref{table:user_dependent}.

The classification model implemented using $U_M$ mode on $D_u$ with a randomized 75\%-25\% train-test split, provides an accuracy of 97.85\%, and its confusion matrix can be showcased in Table \ref{table:uwave_mixed_confmat}.

\begin{table}[ht]
\centering
\caption{Confusion Matrix for $U_M$ mode on $D_u$; Brown: Gesture Signed, Yellow: Gesture Classified, Green: Correct Classifications, Red: Incorrect Classifications}
\label{table:uwave_mixed_confmat}
\begin{tabular}{|c|c|c|c|c|c|c|c|c|}
\hline
\rowcolor[HTML]{FFFE65}
{\color[HTML]{EDB13D} }   & 1                             & 2                             & 3                             & 4                             & 5                             & 6                            & 7                            & 8                             \\ \hline
\cellcolor[HTML]{EDB13D}1 & \cellcolor[HTML]{67FD9A}99.29 & \cellcolor[HTML]{F8524D}0.71  & 0                             & 0                             & 0                             & 0                            & 0                            & 0                             \\ \hline
\cellcolor[HTML]{EDB13D}2 & \cellcolor[HTML]{F8524D}0.71  & \cellcolor[HTML]{67FD9A}99.29 & 0                             & 0                             & 0                             & 0                            & 0                            & 0                             \\ \hline
\cellcolor[HTML]{EDB13D}3 & \cellcolor[HTML]{F8524D}0.71  & 0                             & \cellcolor[HTML]{67FD9A}97.86 & \cellcolor[HTML]{F8524D}1.43  & 0                             & 0                            & 0                            & 0                             \\ \hline
\cellcolor[HTML]{EDB13D}4 & 0                             & 0                             & \cellcolor[HTML]{F8524D}2.14  & \cellcolor[HTML]{67FD9A}97.86 & 0                             & 0                            & 0                            & 0                             \\ \hline
\cellcolor[HTML]{EDB13D}5 & 0                             & 0                             & 0                             & 0                             & \cellcolor[HTML]{67FD9A}97.14 & \cellcolor[HTML]{F8524D}2.86 & 0                            & 0                             \\ \hline
\cellcolor[HTML]{EDB13D}6 & 0                             & 0                             & 0                             & 0                             & 0                             & \cellcolor[HTML]{67FD9A}100    & 0                            & 0                             \\ \hline
\cellcolor[HTML]{EDB13D}7 & \cellcolor[HTML]{F8524D}1.43  & \cellcolor[HTML]{F8524D}1.43                          & 0                             & 0                             & 0                             & 0                            & \cellcolor[HTML]{67FD9A}95   & \cellcolor[HTML]{F8524D}3.57  \\ \hline
\cellcolor[HTML]{EDB13D}8 & 0                             & \cellcolor[HTML]{F8524D}0.71  & 0                             & 0                             & 0                             & 0                            & \cellcolor[HTML]{F8524D}2.86 & \cellcolor[HTML]{67FD9A}96.43 \\ \hline
\end{tabular}
\end{table}

It was seen that, on applying $U_I$ mode on $D_u$, an average efficiency (average Leave-One-User-Out Cross-Validation score) of 82.49\% was observed, while \cite{cite:uwave} has achieved 75.4\%. The confusion matrix for the last user as test, trained upon the first seven users, which yields an accuracy of 92.14\%, has been illustrated in Table \ref{table:uwave_independent_confmat}.

\begin{table}[ht]
\centering
\caption{Confusion Matrix for $U_I$ mode on $D_u$; Brown: Gesture Signed, Yellow: Gesture Classified, Green: Correct Classifications, Red: Incorrect Classifications}
\label{table:uwave_independent_confmat}
\begin{tabular}{|c|c|c|c|c|c|c|c|c|}
\hline
\rowcolor[HTML]{FFFE65}
{\color[HTML]{000000} }   & 1                             & 2                            & 3                             & 4                             & 5                         & 6                             & 7                            & 8                             \\ \hline
\cellcolor[HTML]{EDB13D}1 & \cellcolor[HTML]{67FD9A}95.71 & \cellcolor[HTML]{F8524D}4.29 & 0                             & 0                             & 0                         & 0                             & 0                            & 0                             \\ \hline
\cellcolor[HTML]{EDB13D}2 & 0                             & \cellcolor[HTML]{67FD9A}100    & 0                             & 0                             & 0                         & 0                             & 0                            & 0                             \\ \hline
\cellcolor[HTML]{EDB13D}3 & 0                             & 0                            & \cellcolor[HTML]{67FD9A}82.86 & \cellcolor[HTML]{F8524D}17.14 & 0                         & 0                             & 0                            & 0                             \\ \hline
\cellcolor[HTML]{EDB13D}4 & 0                             & 0                            & 0                             & \cellcolor[HTML]{67FD9A}100     & 0                         & 0                             & 0                            & 0                             \\ \hline
\cellcolor[HTML]{EDB13D}5 & 0                             & 0                            & 0                             & 0                             & \cellcolor[HTML]{67FD9A}100 & 0                             & 0                            & 0                             \\ \hline
\cellcolor[HTML]{EDB13D}6 & 0                             & \cellcolor[HTML]{F8524D}4.29 & 0                             & 0                             & 0                         & \cellcolor[HTML]{67FD9A}95.71 & 0                            & 0                             \\ \hline
\cellcolor[HTML]{EDB13D}7 & 0                             & \cellcolor[HTML]{F8524D}1.43 & 0                             & 0                             & 0                         & 0                             & \cellcolor[HTML]{67FD9A}70   & \cellcolor[HTML]{F8524D}28.57 \\ \hline
\cellcolor[HTML]{EDB13D}8 & 0                             & \cellcolor[HTML]{F8524D}1.43 & 0                             & 0                             & 0                         & 0                             & \cellcolor[HTML]{F8524D}5.71 & \cellcolor[HTML]{67FD9A}92.86 \\ \hline
\end{tabular}
\end{table}

Table \ref{table:uwave_eff_time} shows that $RC$ has the least classification time along with an accuracy of 97.5\%, which is marginally lesser than $ET$, thereby effectively capturing the similarities between gesture samples signed by the same user, i.e, in $U_D$ mode. $GB$ provided accuracies and computational times which are intermediary between $ET$ and $RC$ across all three modes, thereby giving the user a choice to reduce classification time by a significant margin, while having a reasonable accuracy.

\begin{table}[ht]
\centering
\caption{Accuracies (Acc) and Time taken (in seconds) for classification of a single gesture sample 	for $D_u$ using the 3 selected classifiers across all modes}
\label{table:uwave_eff_time}
\begin{tabular}{|c|c|c|c|c|c|c|}
\hline
Classifier$\backslash$Mode   & \multicolumn{2}{c|}{$U_D$} & \multicolumn{2}{c|}{$U_M$} & \multicolumn{2}{c|}{$U_I$} \\ \hline
     & Acc         & Time        & Acc         & Time        & Acc         & Time        \\ \hline
$ET$ & 97.76       & 0.6287      & 97.85       & 0.6873      & 82.49      & 0.6853      \\ \hline
$GB$ & 93.75       & 0.0072      & 94.38       & 0.0076      & 75.64      & 0.0078      \\ \hline
$RC$ & 97.5        & 0.0013      & 83.84       & 0.0013      & 74.64      & 0.0013      \\ \hline
\end{tabular}
\end{table}

\subsection{Sony Dataset Analysis}
\label{subsection:results_C}

For all eight participants, the average accuracy for $D_S$ in $U_D$ mode was observed to be 95.88\% using $ET$, where each user's data was divided into a 75\%-25\% train-test split in random, as done in $D_u$. The efficiencies across all individual users can be observed in Table \ref{table:user_dependent}.

\begin{table}[ht]
\centering
\caption{Accuracies (Acc) and Time taken (in seconds) for classification of a single gesture sample 	for $D_S$ using the 3 selected classifiers across all modes}
\label{table:sony_eff_time}
\begin{tabular}{|c|c|c|c|c|c|c|}
\hline
Classifier$\backslash$Mode   & \multicolumn{2}{c|}{$U_D$} & \multicolumn{2}{c|}{$U_M$} & \multicolumn{2}{c|}{$U_I$} \\ \hline
     & Acc         & Time        & Acc         & Time        & Acc         & Time        \\ \hline
$ET$ & 95.88      & 0.6038      & 98.63      & 0.6538      & 75.1      & 0.669       \\ \hline
$GB$ & 90.5        & 0.0055      & 95.5        & 0.0057      & 66.41       & 0.0057      \\ \hline
$RC$ & 94.13      & 0.0013      & 77.75       & 0.0014      & 61.59       & 0.0013      \\ \hline
\end{tabular}
\end{table}

\begin{table*}[ht]
\centering
\caption{Best Accuracies (Acc) and Least Times taken (in seconds) for classification of a single gesture sample across all modes for both datasets $D_u$ and $D_S$}
\label{table:final}
\begin{tabular}{|c|c|c|c|c|c|c|}
\hline
\multirow{2}{*}{Dataset$\backslash$Mode} & \multicolumn{2}{c|}{$U_D$}            & \multicolumn{2}{c|}{$U_M$}                & \multicolumn{2}{c|}{$U_I$}          \\ \cline{2-7} 
                           & Acc ($ET$) & Time ($RC$) & Acc ($ET$) & Time ($RC$) & Acc ($ET$) & Time ($RC$)\\ \hline
$D_u$                     & 97.76                & 0.0013                   & 97.85                & 0.0013                  & 82.49                & 0.0013                   \\ \hline
$D_S$                     & 95.88                & 0.0013                   & 98.63               & 0.0014                  & 75.1                 & 0.0013                   \\ \hline
\end{tabular}
\end{table*}

In the $U_M$ mode, the proposed model with a 75\%-25\% train-test split at random on $D_S$, yields an efficiency of 98.625\%. Table \ref{table:sony_mixed_confmat} shows the confusion matrix for the $U_M$ mode across all 20 gestures.

An average 8-fold cross validation score (average recognition rate across all users) of 75.093\% was observed in the $U_I$ mode of $D_S$. The confusion matrix trained upon the first seven users with the last user as test, which yields an accuracy of 91.75\%, is shown in Table \ref{table:sony_independent_confmat}.

\begin{figure}[ht]
\centering
\includegraphics[width=\linewidth, height=170pt]{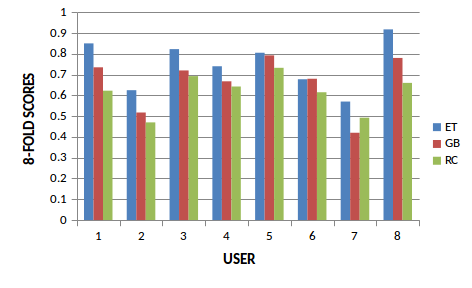}
\caption{8-fold Cross-validation scores for all users in $D_S$}
\label{fig:sony_27usersgraph}
\end{figure}

Figure \ref{fig:sony_27usersgraph} shows the behavior of all users as test (Leave-One-User-Out Cross-Validation scores) in the $U_I$ mode on $D_S$. The large number of misclassifications in the second and seventh user is indicative of both users having not performed the gestures in a manner similar to the rest, thereby reducing the overall efficiency across all classifiers.

Owing to the generic FeatureSet ($FS$) implemented, the results observed in Table \ref{table:sony_eff_time} for $D_S$ are analogous to the results observed in $D_u$ for all three chosen classifiers across all the three modes. 

\section{Conclusion}
\label{section:conclusion}

A Gesture Recognition system with the capability to operate in any of the three modes - User Dependent, Mixed User and User Independent, with a set of generic features was designed, validated and tested in this paper. The proposed system was tested on two public accelerometer-based gesture datasets - uWave and Sony. Table \ref{table:final} showcases the best classifier for each category - Efficiency and Classification Time for a gesture sample, across all three modes of both the datasets. As can be seen in Table \ref{table:final}, Extremely Randomized Trees was observed to perform the best in terms of accuracy, while Ridge Classifier provided the least classification time which was around 500 times faster than $ET$ for a gesture sample, irrespective of the mode or dataset. 

The end-users are given the flexibility to choose any combination of modes and classifiers according to their requirements. This system was implemented on a Raspberry Pi Zero priced at 5 USD making it a low-cost alternative.

\section{Acknowledment}

The authors would like to thank Solarillion Foundation for it’s support and funding of the research work carried out.

\end{document}